\input{aipcheck}

\documentclass[final]{aipproc}
\usepackage{amsmath}
\usepackage[utf8]{inputenc}
\usepackage[ngerman,english]{babel}
\newcommand{\tr}{\operatorname{Tr}}
\layoutstyle{8x11single}
\begin{document}

\hfill {\bf ITEP-LAT-2014-16, HU-EP-14/52}

\title{Two-Color QCD with Chiral Chemical Potential
\footnote{This paper is based on talks given by A. Yu. Kotov 
at Conferences `LATTICE '14', New York, 2014 and `QCHS XI', 
St. Petersburg, 2014.}}

\classification{12.38.Aw, 12.38.Gc, 11.15.Ha}
\keywords{Lattice gauge theory, deconfinement, 
 chiral symmetry breaking, phase transition, chiral density}

\author{V. V. Braguta}{
  address={Institute of Theoretical and Experimental Physics, 
  117218 Moscow, Russia}
}
\author{V. A. Goy}{
  address={Far Eastern Federal University, School of Biomedicine, 
  690950 Vladivostok, Russia}
}
\author{E.-M. Ilgenfritz}{
  address={Joint Institute for Nuclear Research, VBLHEP and BLTP, 
  141980 Dubna, Russia},
}
\author{A.~Yu.~Kotov}{
  address={Institute of Theoretical and Experimental Physics, 
  117218 Moscow, Russia}
}
\author{A. V. Molochkov}{
  address={Far Eastern Federal University, School of Biomedicine, 
  690950 Vladivostok, Russia}
}
\author{M.~M\"uller-Preussker}{
  address={Humboldt-Universit\"at zu Berlin, Institut f\"ur Physik, 
  12489 Berlin, Germany}
}
\author{B.~Petersson}{
  address={Humboldt-Universit\"at zu Berlin, Institut f\"ur Physik, 
  12489 Berlin, Germany}
}
\author{A.~Schreiber}{
  address={Humboldt-Universit\"at zu Berlin, Institut f\"ur Physik, 
  12489 Berlin, Germany}
}

\begin{abstract}
The phase diagram of two-color QCD with a chiral chemical potential 
is studied on the lattice. The focus is on the confinement/deconfinement 
phase transition and the breaking/restoration of chiral symmetry. The 
simulations are carried out with dynamical staggered fermions without 
rooting. The dependence of the Polyakov loop, the chiral condensate and 
the corresponding susceptibilities on the chiral chemical potential and 
the temperature are presented.
\end{abstract}

\maketitle

\section{Introduction}

For the vacuum state of QCD and the properties of low-temperature QCD
the existence of non-trivial topological excitations is important.
Well known are instantons~\cite{Belavin:1975fg} as classical solutions 
in Euclidean space. The role of topology for the solution of the famous 
$U_A(1)$ problem has been recognized very early~\cite{Witten:1979vv,
Veneziano:1979ec}.

It is known by now from lattice QCD that the (fractal) dimensionality of 
the topological structures in the vacuum depends very strongly on the 
resolution scale~\cite{Ilgenfritz:2008ia}. 
In particular, infrared instantons structures are believed to explain 
chiral symmetry breaking~\cite{Schafer:1996wv,Shuryak:1997qb}.

The gluon fields contributing to the path integral at finite temperature
correspondingly may contain calorons~\cite{Kraan:1998pm,Lee:1998bb}.
Because of non-trivial holonomy they consist of dyons and therefore,    
have a richer structure than instantons. Their changes they experience 
at the QCD phase transition are presently under 
study~\cite{Ilgenfritz:2013oda,Bornyakov:2014esa}.

Some time ago the gluonic topological structure and the famous axial 
anomaly have been proposed to be immediately observable (and controllable) 
through the generation of $P$ and $CP$ violating domains in heavy ion 
collisions~\cite{Kharzeev:2004ey,Fukushima:2008xe}. 
It has been demonstrated by detailed numerical calculations
\cite{Kharzeev:2004ey,Kharzeev:2001ev} that macroscopic domains 
of (anti)parallel color-electric and color-magnetic field can emerge in 
a heavy ion collision creating an increasing chiral imbalance 
among the quarks which are deconfined due to the high temperature.
In this situation, the magnetic field created by the spectator nucleons
may initiate a charge separation relative to the reaction plane (parallel
to the electro-magnetic field)~\cite{Kharzeev:2007jp}. 
The resulting charge asymmetry of quarks would become observable in 
terms of recombined hadrons (chiral-magnetic effect)
\cite{Abelev:2009ac,Abelev:2009ad}.
The strength (and particularly the dependence on the 
collision energy) of this effect has been theoretically studied and 
proposed to be a measure for the transient existence of liberated 
quarks~\cite{Kharzeev:2004ey,Fukushima:2008xe,Adamczyk:2014mzf}.

In recent years the dependence of the chiral and deconfinement transitions
on the magnetic field has been investigated both in models and ab-initio
lattice simulations, see e.g. ~\cite{Shovkovy:2012zn,D'Elia:2012tr}. It 
remains an open question whether the phase transition from quarks to hadrons,
i.e. the onset of confinement and chiral symmetry breaking (and vice versa), 
depends on the chiral imbalance.

In this article we study the change of the phase structure by an equilibrium 
lattice simulation. We mimic the topological content (of a topologically 
nontrivial gluonic background in heavy ion collisions) by a standard of chiral 
imbalance, which is provided by a chiral chemical condensate. In this form, 
the modification of the phase diagram by the chiral chemical potential $\mu_5$
has been studied mainly in effective models \cite{Fukushima:2010fe,
Chernodub:2011fr,Gatto:2011wc,Andrianov:2013dta,Andrianov:2013qta} 
with which we will compare our results.

On the lattice, contrary to the ordinary chemical potential (for quarks, i.e. 
baryonic charge), simulations with non-zero $\mu_5$ are not hampered by a 
sign problem. They are accessible to standard hybrid Monte Carlo algorithms. 
Such lattice simulations with $\mu_5 \ne 0$ were already performed in 
Ref.~\cite{Yamamoto:2011gk,Yamamoto:2011ks}. The main goal of these papers, 
however, was the chiral magnetic effect. Therefore, the phase diagram was 
not systematically studied.    

In our study we perform simulations with the $SU(2)$ gauge group.  
One reason is that less computational resources are required for 
this pilot study than for full QCD. The second one is that we have already
carried out two-colour QCD computations with an external magnetic field 
~\cite{Ilgenfritz:2012fw,Ilgenfritz:2013ara}.

\section{Details of the simulations}

We have performed simulations with the $SU(2)$ gauge group. We employ 
the standard Wilson plaquette action
\begin{equation}\begin{split}
S_g = \beta\sum\limits_{x,\mu<\nu}&\left(1-
       \frac{1}{N_c}\tr U_{\mu\nu}(x)\right).
\end{split}\end{equation}
For the fermionic part of the action we use staggered fermions 
\begin{equation}\begin{split}
S_f=ma\sum_x {\bar \psi_x}\psi_x+&\frac12\sum_{x\mu} 
   \eta_{\mu}(x)({\bar \psi_{x+\hat{\mu}}U_{\mu}(x)}\psi_x
   -{\bar \psi_x}U^{\dag}_{\mu}(x) \psi_{x+\hat{\mu}})+\\
   +&\frac12\mu_5a
   \sum_x s(x)({\bar \psi}_{x+\hat{\delta}}{\bar U}_{x+\hat{\delta}, x}\psi_x
   -{\bar \psi}_{x}{\bar U}_{x+\hat{\delta}, x}^{\dag}\psi_{x+\hat{\delta}}),
\label{eq:staggeredaction}
\end{split}\end{equation}
where the $\eta_{\mu}(x)$ are the standard staggered phase factors: 
$\eta_1(x)=1,\eta_{\mu}(x)=(-1)^{x_1+\ldots+x_{\mu-1}}$ for $\mu=2,3,4$. 
Furthermore, $a$ denotes the lattice spacing, $m$ the bare fermion mass, and 
$\mu_5$  the value of the chiral chemical potential. In the chirality 
breaking term $s(x)=(-1)^{x_2}$, $\delta=(1,1,1,0)$ represents a shift to 
a diagonally located site of a spatial elementary cube. The combination
${\bar U}_{x+\hat{\delta},x}=\frac16\sum\limits_{i,j,k=
\text{perm}(1,2,3)}U_i(x+\hat{e}_j+\hat{e}_k)U_j(x+\hat{e}_k)U_k(x)$ is connecting sites 
$x$ and $x+\hat{\delta}$ symmetrized over the $6$ shortest paths between these 
sites. In the continuum limit Eq. (\ref{eq:staggeredaction}) can be 
rewritten in the Dirac spinor-flavor basis 
\cite{KlubergStern:1983dg,MontvayMuenster:2000} as follows
\begin{equation}\begin{split}
S_f \to S^{(cont)}_f = \int d^4x \sum_{i=1}^4 \bar{q_i}
 (\partial_{\mu}\gamma_{\mu} + 
 igA_{\mu}\gamma_{\mu}+m+\mu_5\gamma_5\gamma_4)q_i.
\end{split}\end{equation}

It should be noted here that the usual baryonic chemical potential 
\cite{Hasenfratz:1983ba} and also the chiral chemical potential 
\cite{Yamamoto:2011ks} are introduced to the action as a 
modification of the temporal links by corresponding exponential factors in 
order to eliminate chemical potential dependent quadratic divergencies.
For staggered fermions this modification can be performed as well for the 
baryonic chemical potential. However, for the chiral chemical potential such 
a modification leads to a highly non-local action \cite{Yamamoto:2011ks}. 
Therefore, we decided to introduce $\mu_5$ in Eq.~(\ref{eq:staggeredaction}) 
in an additive way similar to the mass term leaving aside the question of 
arising singularities for the time being. We expect that the Polyakov loop 
will not contain $\mu_5$ dependent singular terms. 

We have performed simulations with two lattice sizes 
$N_{\tau}\times N_{\sigma}^3=6\times16^3, 10\times28^3$. 
The measured observables are
\begin{itemize}
\item the Polyakov loop 
\begin{equation}\begin{split}
L=\frac{1}{N_\sigma^3}\sum_{n_1,n_2,n_3}
\langle\tr\prod\limits_{n_4=1}^{N_{\tau}}U_4(n_1,n_2,n_3,n_4)\rangle \,,
\end{split}\end{equation}
\item the chiral condensate
\begin{equation}\begin{split}
a^3\langle\bar{\psi}\psi\rangle  = 
-\frac{1}{N_{\tau}N_{\sigma}^3}\frac14\frac{\partial}{\partial(ma)}\log Z = 
\frac{1}{N_{\tau}N_{\sigma}^3}\frac14\langle \tr \frac{1}{D+ma} \rangle \,,
\end{split}\end{equation}
\item the Polyakov loop susceptibility
\begin{equation}\begin{split}
\chi_L=N_{\sigma}^3\left(\langle L^2\rangle-\langle L\rangle^2\right)\,,
\end{split}\end{equation}
\item the disconnected part of the chiral susceptibility
\begin{equation}\begin{split}
\chi_{disc}=\frac{1}{N_{\tau}N_{\sigma}^3}\frac1{16}
 (\langle (\tr \frac{1}{D+ma})^2 \rangle-
 \langle \tr \frac{1}{D+ma} \rangle^2)\,.
\end{split}\end{equation}
\end{itemize} 
The Polyakov loop and the corresponding susceptibility are sensitive to 
the confinement/deconfinement phase transition, while the chiral condensate 
in principle responds to chiral symmetry breaking/restoration.

The simulations have been carried out with a CUDA code to run the Hybrid 
Monte Carlo algorithm on GPU's. The dependence of the lattice spacing on 
the coupling parameter $\beta$ was taken from \cite{Ilgenfritz:2012fw}. 
For our simulations with the lattice size $6\times16^3$ the fermion mass 
was kept fixed in lattice units at $ma=0.01$ while changing $\beta$. 
E.g. for $\beta=1.80$ this corresponds to a pion mass value 
$m_{\pi} \approx 330~\text{MeV}$. 
For the larger lattice size $10\times28^3$ and for various $\beta$ values 
we have chosen the same bare quark mass in physical units 
$m \simeq 19~\mathrm{MeV}$, which corresponds to 
$m_{\pi} \approx 540~\text{MeV}$.

\section{Results and conclusions}

\begin{figure}[!th]
\begin{tabular}{cc}
\includegraphics[scale=0.60,clip=false]{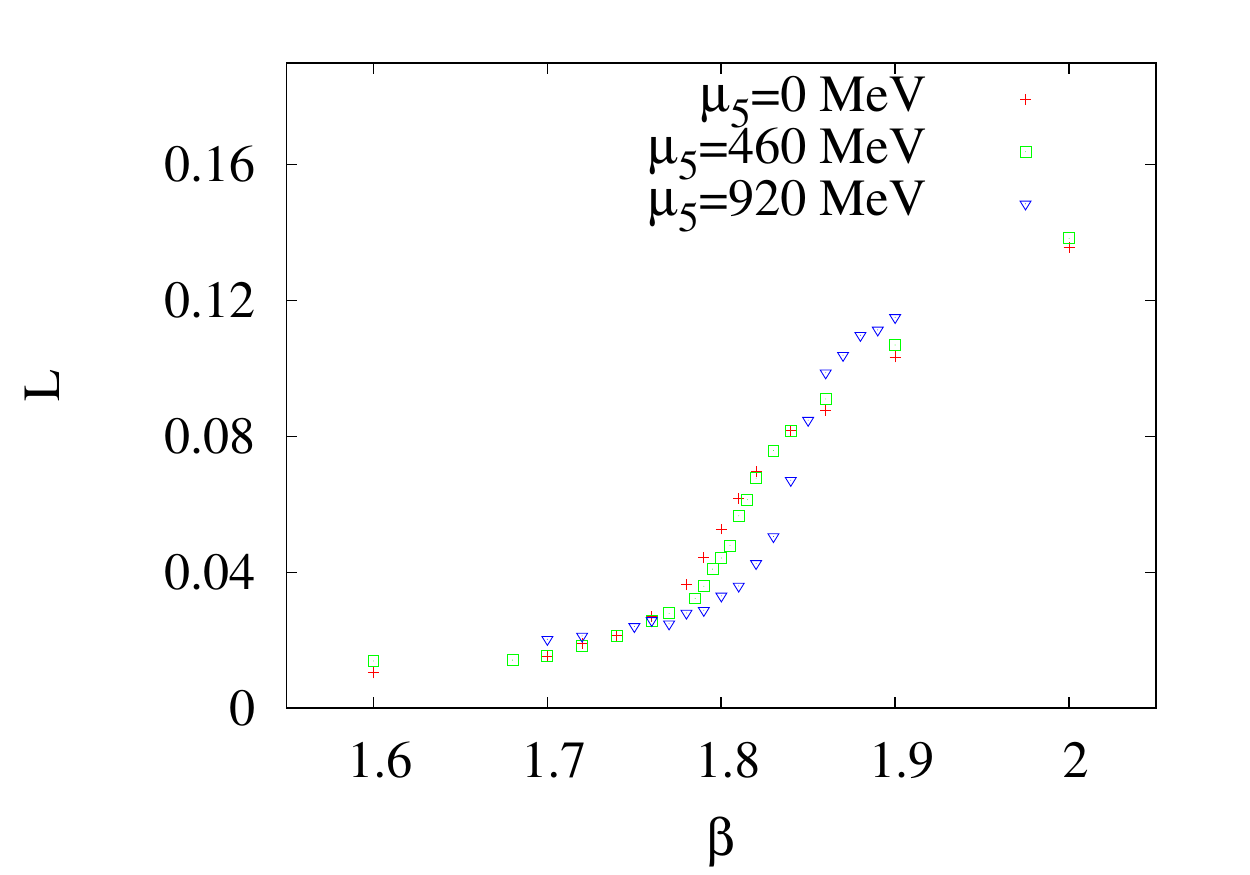} & 
\includegraphics[scale=0.60,clip=false]{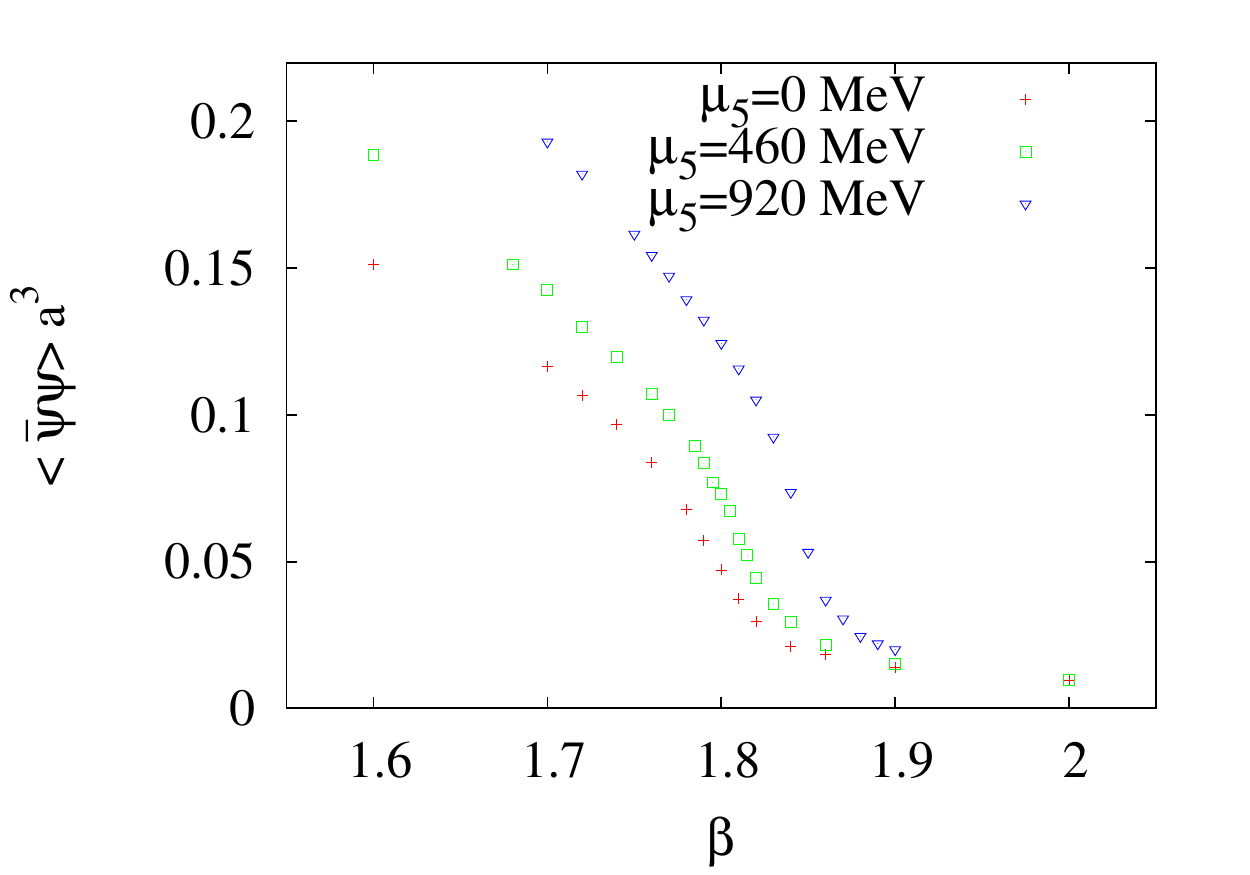} 
\end{tabular}
\caption{Polyakov loop (left) and chiral condensate (right)
versus $\beta$ for three $\mu_5$ values and lattice size 
$6\times16^3$. Errors are smaller than the data point symbols. }
\label{fig:obs}
\end{figure}

\begin{figure}[!hb]
\begin{tabular}{cc}
\includegraphics[scale=0.60,clip=false]{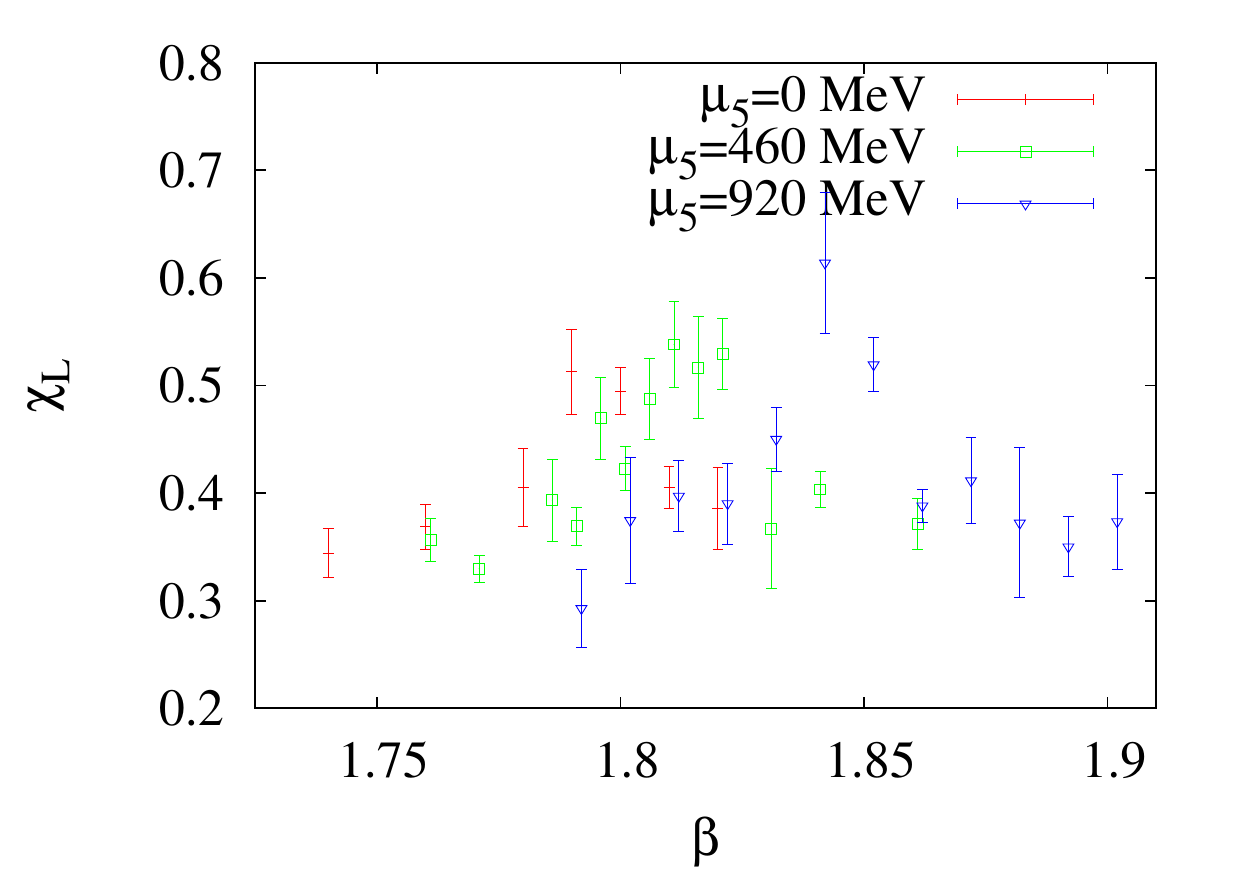} & 
\includegraphics[scale=0.60,clip=false]{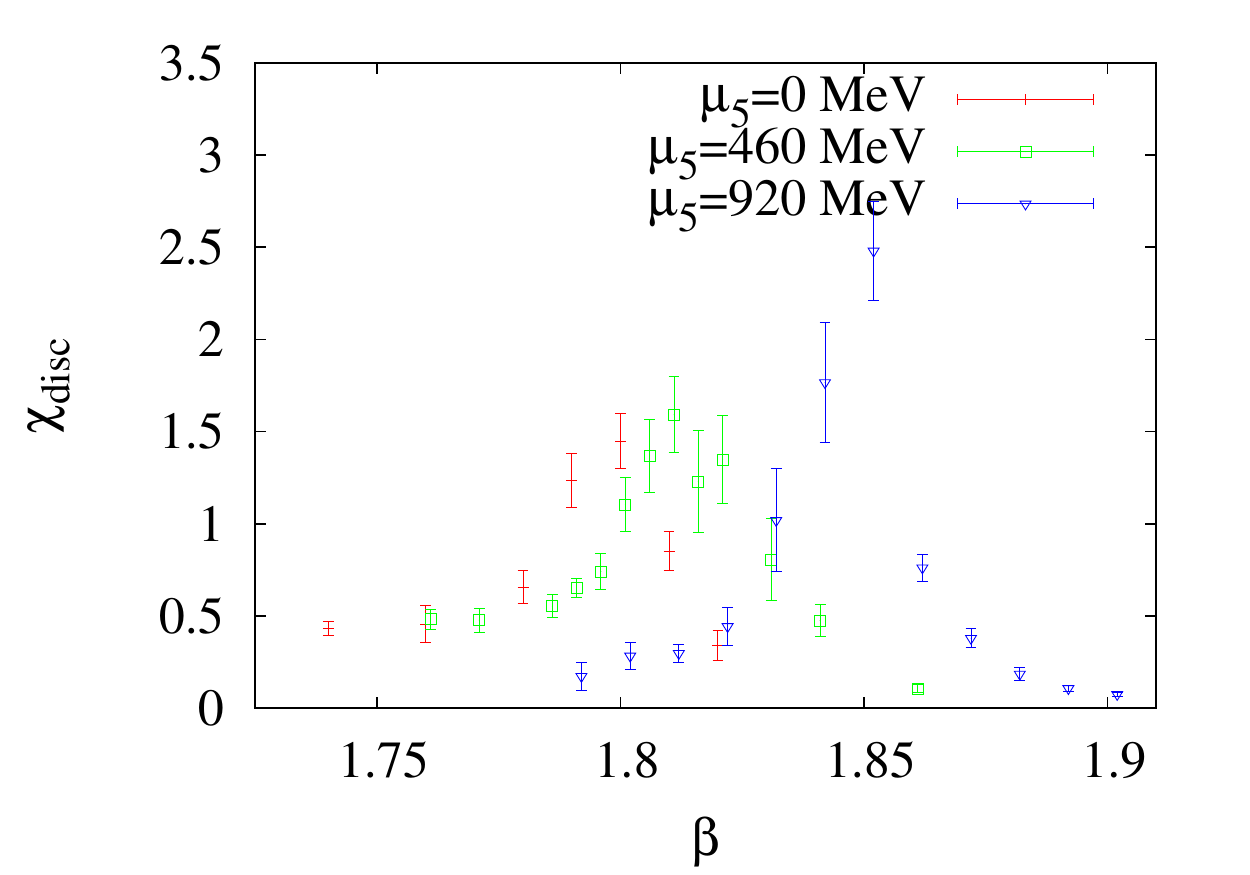} 
\end{tabular}
\caption{Polyakov loop susceptibility (left) and chiral 
susceptibility (right) versus $\beta$ for three values of 
$\mu_5$ and lattice size $6 \times 16^3$. In order to avoid a 
complete superposition of data points belonging to 
different $\mu_5$ values we applied a tiny shift along 
the $\beta$ axis.\vspace*{0.5cm}}
\label{fig:sus}
\end{figure}

For the smaller lattice ($6 \times 16^3)$ we present results for three 
fixed values of $\mu_5=0, 460, 920$ MeV and for different values of 
$\beta$, while the bare fermion mass remained constant in lattice units 
$ma=0.01$. The results for the Polyakov loop and the chiral condensate 
are plotted in Fig. \ref{fig:obs}. We see that increasing chiral chemical 
potential moves the position of the deconfinement and chiral transition, 
respectively, to larger values of $\beta$.  This means that the transition 
temperature increases. Plots for the chiral susceptibility and the Polyakov 
loop susceptibility (see Fig. \ref{fig:sus}) confirm this observation. 
We estimate the change of the critical temperature to be 
$\frac{T_c(\mu_5)-T_c(0)}{T_c(0)}\sim 20\%$ for $\mu_5=920$ MeV. 
The results do not show any splitting between the chiral and the 
deconfinement transition.

\vspace*{0.5cm}
\begin{figure}[!th]
\begin{tabular}{cc}
\includegraphics[scale=0.60,clip=false]{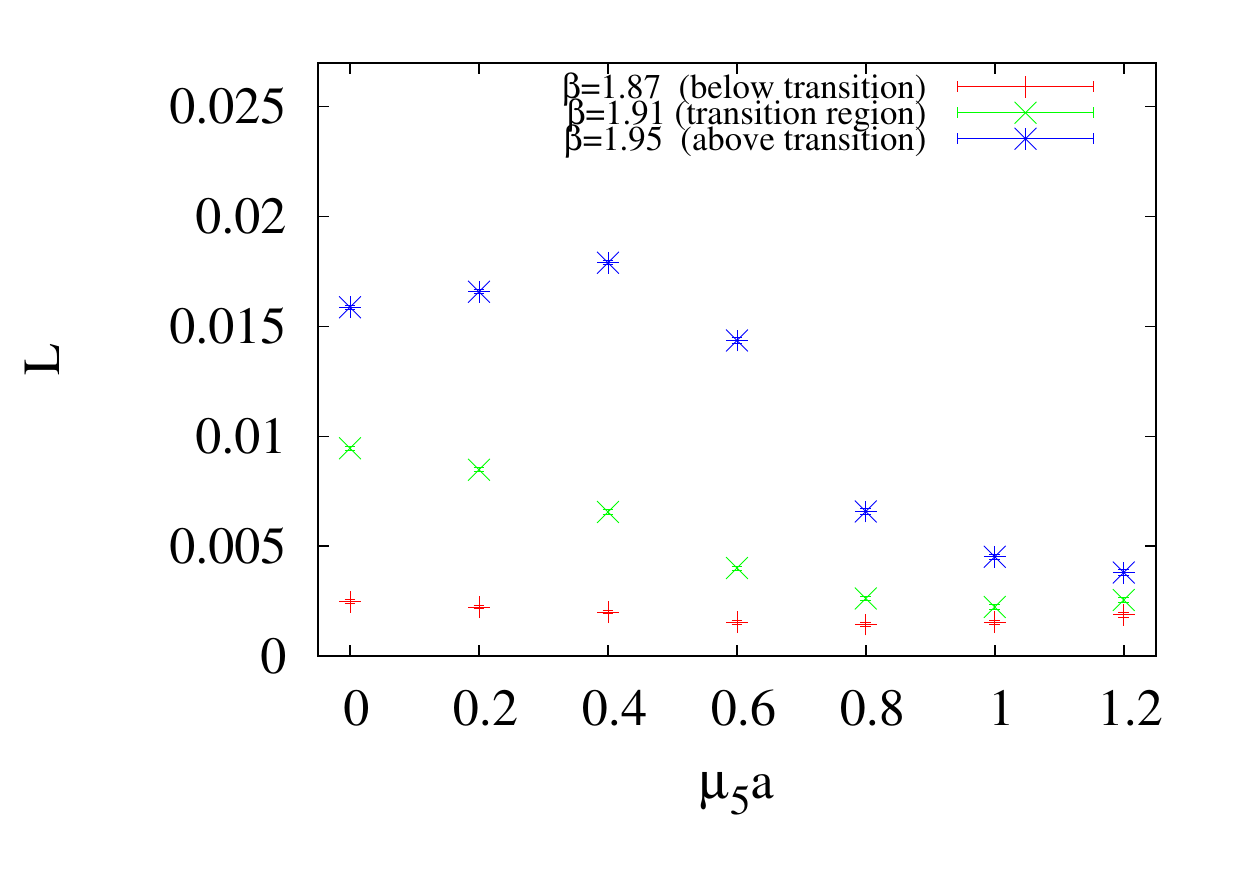} & 
\includegraphics[scale=0.60,clip=false]{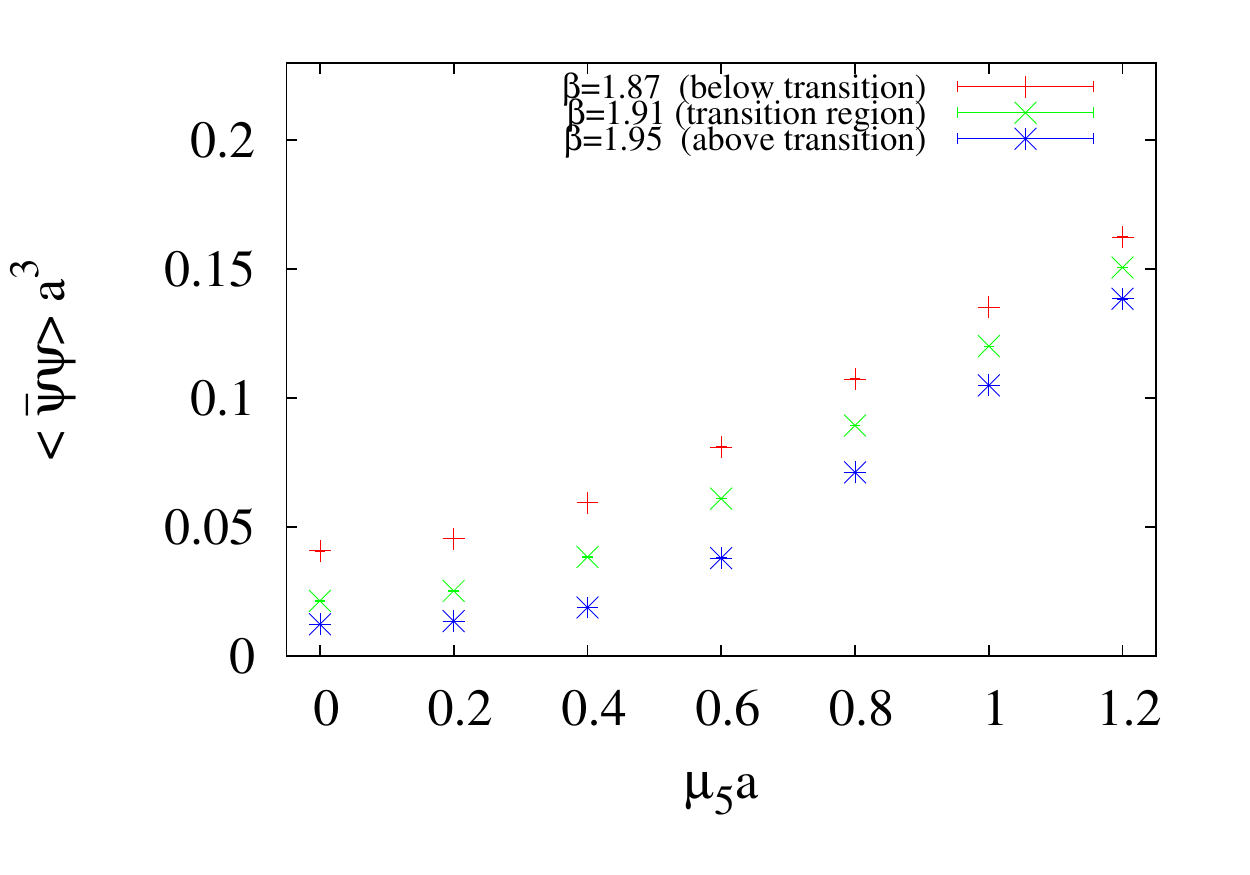} 
\end{tabular}
\caption{Polyakov loop (left) and chiral condensate (right) 
versus $\mu_5 a$ for three fixed values of $\beta$ and 
lattice size $10 \times 28^3$.\vspace*{0.5cm}}
\label{fig:largeobs}
\end{figure}

To confirm our results we carried out simulations also for the larger 
lattice size $10\times28^3$. In Fig. \ref{fig:largeobs} we present results 
for varying $\mu_5$  and fixed values of $\beta=1.87, 1.91$ and $1.95$, 
which for a vanishing chiral chemical potential correspond to temperatures 
below the transition, to the transition region, and to the high temperature 
phase, respectively. As can be seen from this figure, in the confinement 
phase the Polyakov loop remains almost constant with increasing chiral 
chemical potential. It means, that if the system was in the confinement 
phase at $\mu_5 =0$, it remains confined at $\mu_5 > 0$. Moreover, we 
observe the Polyakov loop to drop down both in the deconfinement phase and 
in the transition region. Thus, the system goes into the confinement phase 
for sufficiently large $\mu_5$. With other  words, we conclude that the 
critical temperature increases with an increasing chiral chemical potential 
in agreement with our results obtained on the smaller lattice. Notice 
that in case of the larger lattice we have kept fixed the bare 
fermion mass in physical units at $m \simeq 19~\mathrm{MeV}$ for all 
three $\beta$ values, while changing $\mu_5$. It is worth mentioning 
that the behavior described above looks quite similar as that obtained for 
two-color QCD in an external magnetic field 
\cite{Ilgenfritz:2012fw,Ilgenfritz:2013oda}.

Our results are in contradiction to those of the models studied in 
\cite{Fukushima:2010fe,Chernodub:2011fr,Gatto:2011wc}, where the critical 
temperature of the transition was observed to decrease. Furthermore, in 
these papers at some critical value of the chiral chemical potential the 
transition was reported to become first order. In our simulations we do 
not see such a behavior. However, the results obtained have a tendency 
towards a sharper phase transition at nonzero chiral chemical potential. 

Although the analytic results are only derived in models and not in full 
QCD also in our approach there are some differences to QCD. We use the 
$SU(2)$ gauge group instead of $SU(3)$ and four flavor degrees of freedom. 
Moreover, the pion mass value used here is higher than the physical one. 
The situation can change, when one arrives at smaller quark masses. 
We want to address this question in a future work.

\begin{theacknowledgments}
The authors are grateful to V. I. Zakharov and V. G. Bornyakov for 
interesting and stimulating discussions. The simulations were performed 
at GPUs of supercomputer K100 and computers of the Berlin group. 
The work was supported by FEFU grants No. 12-02-13000-FEFU{\_}a 
and 13-09-617-m{\_}a, RFBR grants 14-02-01185-a, 13-02-01387-a, 
grant of the president of the RF MD-3215.2014.2 and grant of the 
FAIR-Russia Research Center.
\end{theacknowledgments}

\bibliographystyle{aipproc}   

\end{document}